\documentclass[a4paper,11pt]{article}
\usepackage{pos}
\newcommand{\be}{\begin{equation*}}
\newcommand{\ee}{\end{equation*}}
\newcommand{\bea}{\begin{eqnarray*}}
\newcommand{\eea}{\end{eqnarray*}}

\def\lQ{\Lambda_{\rm QCD}}

\newcommand{\nn}{\nonumber}
\newcommand{\al}{\alpha}

\newcommand{\MS}{\overline{\rm MS}}

\newcommand{\PV}{\rm PV}

\usepackage{amsmath}

\title{Hyperasymptotic Expansions in QCD: The Lightest Gluelump Mass Case}

\author*[a,b]{Antonio Pineda}

\affiliation[a]{Grup de Física Teòrica, Dept. Física, Universitat Autònoma de Barcelona,\\
  E-08193 Bellaterra, Barcelona, Spain}

\affiliation[b]{Institut de Física d'Altes Energies (IFAE), The Barcelona Institute of Science and Technology,\\
Campus UAB, 08193 Bellaterra (Barcelona), Spain}

\emailAdd{AntonioMiguel.Pineda@uab.es} 

\abstract{In these proceedings, we provide a summary of our recent advancements in determining the mass of the lightest gluelump using hyperasymptotic expansions in Quantum Chromodynamics (QCD) \cite{Ayala2025}. This paper details our methodology, our precise determination of leading renormalon normalization constants, and our extraction of the gluelump mass from two independent physical systems, yielding a final combined, renormalization-group-invariant and scheme-independent result of $\Lambda_{B}^{\rm PV} = 2.44(7) r_0^{-1}$.}

\FullConference{Loops and Legs in Quantum Field Theory (LL2026)\\
12-17, April, 2026\\
Bayreuth, Germany\\}

\begin{document}
\maketitle

\section{Introduction}

The study of heavy quarkonium hybrids and gluelumps is driven by a convergence of significant phenomenological and theoretical interests. 

Phenomenologically, the past two decades have seen a rapid proliferation of exotic hadron discoveries (the so-called XYZ states) at facilities like Belle, BaBar, BESIII, and LHCb  \cite{Brambilla:2019esw}. Understanding the internal structure of these states necessitates rigorous theoretical frameworks, as a substantial fraction of these exotics may be interpreted as heavy quarkonium hybrids---mesons where the gluonic degrees of freedom are explicitly excited.

Theoretically, these systems provide a unique window into the behavior of QCD interactions in color representations other than the fundamental one. A significant advantage over studying pure glueballs is that gluelumps possess a static color source, which heavily simplifies the theoretical setup. A fascinating feature of these systems is the intimate relationship between heavy quarkonium hybrids and gluelumps in the short-distance limit (small interquark separation $r$). Within the framework of potential Non-Relativistic QCD (pNRQCD) \cite{Pineda:1997bj,Brambilla:1999xf}, this short-distance regime can be systematically organized in an expansion in powers of $r$. In this limit, the two heavy quarks in an octet configuration appear to the surrounding gluon field as a single, point-like static adjoint source.

Therefore, gluelumps and heavy quarkonium hybrids serve as critical laboratories for understanding (non-)perturbative QCD dynamics beyond the fundamental color representation. Despite the availability of high-order perturbative calculations 
\cite{Bauer:2011ws,Bali:2013pla,Bali:2013qla,Fischler:1977yf,Gorishnii:1991hw,Schroder:1998vy,Brambilla:1999qa,Smirnov:2008pn, Anzai:2009tm, Smirnov:2009fh, Lee:2016cgz,
Kniehl:2004rk,Anzai:2013tja,Pineda:2000gza,Brambilla:2006wp,Brambilla:2009bi,Pineda:2011aw,Pineda:2011db} and precise lattice data \cite{Foster:1998wu,Herr:2023xwg,Bali:2003jq,Schlosser:2021wnr}, the asymptotic nature of the perturbative series---driven by renormalon \cite{tHooft} divergences---poses a severe challenge to quantitative Operator Product Expansion (OPE) analyses. By employing a Principal Value (PV) summation scheme alongside systematic hyperasymptotic expansions \cite{Ayala:2019uaw,Ayala:2019hkn}, 
we circumvent these limitations without introducing spurious power-like infrared cutoffs.

\section{The Renormalon Problem and the OPE}
In any OPE analysis, a physical observable depending on a hard scale $Q$ is factorized into short-distance Wilson coefficients and long-distance matrix elements:
\begin{equation}
\label{OPE}
\text{Observable}(Q) = S_{\rm pert}(\alpha_s(Q)) + \sum_d \mathcal{C}_{d}(\alpha_s(Q)) \frac{\langle \mathcal{O}_d \rangle}{Q^d}
\end{equation}
The perturbative series 
$$
S_{\rm pert}(\alpha_s) = \sum_{n=0}^{\infty} p_n \alpha_s^{n+1}
$$
 is divergent. In QCD, the coefficients grow factorially, $p_n \sim K^n \Gamma(n+1 + b)$, driven by the integration over infrared loop momenta. These divergences are known as \textit{renormalons} \cite{tHooft}. This makes it impossible to give meaningful values for the nonperturbative terms until the perturbative sum is regularized.

\subsection{Borel Transform and Ambiguities}
To regularize the perturbative sum, one performs a Borel transform:
\begin{equation}
B[S](t) = \sum_{n=0}^{\infty} \frac{p_n}{n!} t^n
\end{equation}
The physical amplitude is given by the inverse Borel transform:
\begin{equation}
\tilde{S}(\alpha_s) = \int_{0}^{\infty} dt \, e^{-t/\alpha_s}  B[S](t)
\end{equation}
However, infrared renormalons manifest as poles or branch cuts on the positive real $t$-axis. Consequently, the integration contour is ambiguous. We adopt the PV regularization scheme:
\begin{equation}
S_{\rm PV}(\alpha_s) \equiv \int_{0,\rm PV}^{\infty} dt \, e^{-t/\alpha_s} B[S](t)
\end{equation}
It is defined by doing the medium summation of the integral slighlty above and below the real axis. 
This summatial scheme is highly desirable because it can be shown that it is exactly independent of the renormalization scale ($\mu$) and independent of the choice of the strong coupling scheme \cite{Ayala:2019uaw,Takaura:2020byt}.

\subsection{Principal Value Summation and Hyperasymptotics}

Direct computation of $S_{\rm PV}$ to infinite orders is impossible. In practice, we only know the first few orders of the perturbative expansion and have a partial knowledge of the analytic behavior of the closest singularities at the origin of the Borel transform. Approximations must be used. The good point is that the hyperasymptotic expansion makes optimal use of the available information.   It allows for a parametric control of the error and a smooth connection with standard perturbation theory.
\begin{itemize}
\item
At low orders we have {\bf standard perturbation theory} and the error scales as
\begin{equation}
S_{\rm PV}(\alpha_X(Q))-\sum_{n=0}^{N}p^{(X)}_n(\frac{\mu}{Q}) \alpha_X^{n+1}(\mu) \sim {\cal O}(\alpha_X^{N+2})
\end{equation}
where $X$ stands for the renormalization scheme used for $\alpha$. 
 \item \textbf{Superasymptotic Approximation.} \\
 When $N$ becomes large, the coefficient multyplying $\alpha_X^{N+2}$ becomes factorially large. Truncating the perturbative series at its minimal term 
\begin{equation}
\label{eq:NP}
N=N_P \sim |d|\frac{ 2\pi}{\beta_0\alpha_X(\mu)}
\,,
\end{equation}
(where $d$ is the dimension associated to the closest singularity in the Borel plane) 
 minimizes the truncation error. This is the superasymptotic approximation and the error scales as
\begin{equation}
S_{\rm PV}(\alpha_X(Q))-\sum_{n=0}^{N_P}p^{(X)}_n(\frac{\mu}{Q}) \alpha_X^{n+1}(\mu) \sim {\cal O}(e^{-|d|\frac{2 \pi }{\beta_0 \alpha_X(Q)}})
\end{equation}
    \item \textbf{Hyperasymptotic Expansion.} \\
   The superasymptotic approximation yields an error that roughly scales as the nonperturbative constants we want to determine.
    To achieve true exponential precision, one introduces "terminants" \cite{BerryandHowls, Dingle}. The generalization of this concept to QCD was worked out in Refs. \cite{Ayala:2019uaw, Ayala:2019hkn}, where general expressions for the terminants can be found. The specific expressions used here can be found in Ref. \cite{Ayala2025}.  A terminant effectively resums the divergent tail of the series associated with a specific renormalon pole and scales as
   \begin{equation}
    \Omega_d \sim Z \sqrt{\alpha}e^{-|d|\frac{2\pi}{\beta_0\alpha}}
    \end{equation}
    Overall, we can systematically improve the theoretical expression for $S_{\rm PV}$ by adding terminants and (modified) perturbative expansion to the hyperasymptotic expansion in a systematic way:
\begin{equation}
\label{SPV2final}
S_{\rm PV}(Q)=\sum_{n=0}^{N_P}p^{(X)}_n(\frac{\mu}{Q}) \alpha_X^{n+1}(\mu)
+\Omega(\mu)+\sum_{n=N_P+1}^{N'_P}(p_n-p_n^{(as)})\alpha_X^{n+1}(\mu)+\Omega'(\mu)+\cdots
\,,
\end{equation}
\item
We are then in the position to determine nonperturbative constants assuming the validity of the {\bf Nonperturbative OPE}
\begin{equation}
\label{Obtruncated}
K_X^{\rm (PV)}\alpha_X^{\gamma}(Q)\frac{\Lambda_X^{d}}{Q^d}\left(1+{\cal O}(\alpha_X(Q))\right)=
\left[{\rm Observable}(\frac{Q}{\lQ})
-
S_{\rm PV}(\alpha_X(Q))\right]
+{\cal O}(\frac{\Lambda_X^{d'}}{Q^{d'}})
\,,
\end{equation}
where $d'$ is the dimension of subleading nonperturbative corrections ($d' > d$).
\end{itemize}

\section{Effective Field Theories for Gluelumps}
Gluelump states consist of a static color source in the adjoint (octet) representation bound to a gluonic field such that the overall state is a color singlet. This essentially mirrors the physical picture of a heavy gluino in the static approximation.

\subsection{Heavy Gluino Effective Theory}
Using a Heavy Gluino Effective Theory (HGET), the bound state mass of a gluino and glue can be expanded in inverse powers of the heavy gluino pole mass ($m_{\tilde{g}}$):
\begin{equation}
M_{H,\tilde{G}} = m_{\tilde{g}} + \Lambda_H + \mathcal{O}(1/m_{\tilde{g}})
\end{equation}
This is highly analogous to the expansion of the $B$ meson mass in Heavy Quark Effective Theory (HQET). In both cases, the pole mass ($m_{\tilde{g}}$ or $m_Q$) suffers from a leading $\mathcal{O}(\Lambda_{QCD})$ renormalon ambiguity. To give this expansion quantitative meaning, one must switch to a well-defined short-distance mass scheme, such as the PV mass:
\begin{equation}
M_{H,\tilde{G}} = m_{\tilde{g},\rm PV} + \Lambda_H^{\rm PV} + \mathcal{O}(1/m_{\tilde{g},\rm PV})
\end{equation}
Here, $\Lambda_H^{\rm PV}$ is a universal, non-perturbative constant, free of explicit infrared cutoffs, and independent of the renormalization scale and scheme used for the strong coupling constant.

\section{Dual Extraction of the Gluelump Mass}
Equipped with the PV framework, and the hyperasymptotic expansions, we extract the non-perturbative gluelump energy $\Lambda_B^{\rm PV}$ by applying Eq. (\ref{Obtruncated}) to two different observables. These observables have been accurately computed via lattice simulations in the quenched approximation. 

\subsection{Method 1: Direct Lattice Gluelump Energy}
The first method utilizes direct lattice measurements of the static adjoint source energy \cite{Foster:1998wu,Herr:2023xwg}. The inverse of the lattice spacing $1/a$ plays an analogous role to $Q$ in Eq. (\ref{OPE}). Eq. (\ref{Obtruncated}) in this case reads
\begin{equation}
\Lambda_H^{\rm PV} = \left[ \Lambda_H^L(a) - \delta m_{A,\rm PV}^L(a) \right]+{\cal O}(a^2)
\end{equation}
where $\delta m_{A,PV}^L(a)$ is the self-energy of a static source in the adjoint representation. The coefficients of its perturbative expansion were obtained in Refs. \cite{Bauer:2011ws,Bali:2013pla,Bali:2013qla} up to $\alpha^{20}$. The $\alpha$ dependence of the terminant is formally known to all orders in terms of the perturbative coefficients of the beta function using similar arguments than those used for the pole mass \cite{Beneke:1994rs}. 

Its overall normalization can be computed numerically. The asymptotic behavior of the perturbative coefficients is given by ($d=1$):
\begin{equation}
c_{A,n} \xrightarrow{n \to \infty} c_{A,n}^{\rm (as)} \propto Z_A \left( \frac{\beta_0}{2\pi d} \right)^n \Gamma(n+1+b)
\end{equation}
By matching the known exact coefficients to the asymptotic formula for a variety of observables that share the same renormalon, in Ref. \cite{Ayala2025}, the following highly precise normalization in the $\overline{\text{MS}}$ scheme for $n_f=0$ (quenched approximation) was obtained:
\be
Z_{A}^{\MS} = -1.343(36)
\ee

Finally, one can apply the hyperasymptotic expansion to this observable that reads
\be
\label{LambdabarPV}
\Lambda_{B}^{\PV}=\Lambda_B^L(a)
-\sum_{n=0}^{N_P}\frac{1}{a}c_{A,n}\alpha_L^{n+1}(a)
-\frac{1}{a}\Omega_A(1/a)-\sum_{N_P+1}^{N'=3N_P}\frac{1}{a}[c_{A,n}-c_{A,n}^{\rm (as)}]\alpha_L^{n+1}(a)+{\cal O}(a^2)
\ee
We apply this formula to the lightest gluelump: $H=B$. 
The result should be independent of at which scale $a$ one determines $\Lambda_{B}^{\PV}$. This is indeed beautifully observed in Figs. 8 and 9 in \cite{Ayala2025}. We observe a flat behavior up to effects consistent with the error and with the estimated size of subleading ${\cal O}(a^2)$ effects. Overall, the analysis in Ref. \cite{Ayala2025} yields
\begin{equation}
\Lambda_{B}^{\rm PV} = 2.47(9) r_0^{-1}
\end{equation}

%%%%%%%
\subsection{Method 2: Static Hybrid Potential in the Short-Distance Limit}
The second method exploits the pNRQCD factorization of heavy quarkonium hybrid static energies (such as the $\Pi_u$ and $\Sigma_u^-$ states), where $1/r$ plays an analogous role to $Q$ in Eq. (\ref{OPE}). In the limit of small interquark separation $r \to 0$, the static energy (hybrid potential) $E_H$ reduces to the sum of $2m_Q$, the octet potential, and the gluelump mass, up to corrections that are $\mathcal{O}(r^2)$ suppressed by the multipole expansion. A similar formula holds for the static energy of the singlet state $\Sigma_g^+$, with the qualification that there is not any gluelump mass (as there is no remaining static colour source in the $r \rightarrow 0$ limit). It is convenient to consider the energy difference between the singlet and the hybrid static energy. The reason is that one can get the continuum ($a \rightarrow 0$) limit for such energy difference when one computes them in the lattice and, consequently, one can just focus on the OPE generated by the expansion in powers of $r$. We apply this reasoning to the lightest gluelump (associated to the $\Pi_u$ hybrid state) in the following. Therefore, the analogous of Eq. (\ref{Obtruncated}) to this case reads
\begin{equation}
 \Lambda_B^{\PV} = \left[(E_{\Pi_u}(r)-E_{\Sigma_g^+}(r)) - V_A^{\PV}(r) \right]+ \mathcal{O}(r^2)
\end{equation}
where $V_A=V_o-V_s$ is the difference between the octet and singlet potential. 

The application of the hyperasymptotic expansion in this case reads
\begin{eqnarray}
\nn
\Lambda_B^{\PV}&=&(E_{\Pi_u}(r)-E_{\Sigma_g^+}(r))
-
\sum_{n=0}^{N_P} V^{(A)}_n \al_{\MS}^{n+1}(\nu_s)
-\frac{1}{r}\Omega_{V_A}- \delta V_A^{\rm RG}(r) 
\\
&&
-\sum_{n=N_P+1}^{3N_P/N_{max}} (V^{(A)}_n-V_n^{(A,\rm as)}) \al_{\MS}^{n+1}(\nu_s)
-\delta E^{(2)\PV}_{A,us}(r;\nu_{us})
+o( r^2)
\label{Method2}
\end{eqnarray}
By construction, $V_A^{\PV}$ is a function of a single scale: $1/r$. We have then applied the hyperasymptotic expansion to it accordingly. As we have already mentioned, we can work in the continuum limit. Therefore, there is no obstruction in doing perturbation theory in the $\MS$ scheme. This allows us to obtain $V_A^{\PV}$ with superasymptotic precision using the coefficients computed in Refs. \cite{Fischler:1977yf,Gorishnii:1991hw,Schroder:1998vy,Brambilla:1999qa,Smirnov:2008pn, Anzai:2009tm, Smirnov:2009fh, Lee:2016cgz,Kniehl:2004rk,Anzai:2013tja}. The inclusion of the terminant $\frac{1}{r} \Omega_{V_A}(1/r)$ goes parallel to the inclusion of the terminant $\frac{1}{a}\Omega_A(1/a)$ in Eq. (\ref{LambdabarPV}). It has the same functionality in $\alpha$ and depends on the same normalization constant $Z_A$. Note, however, that now we work in a completely different scheme, $\MS$, compared with the lattice scheme used in the previous section (beware that this affects the coupling but also the normalization). In this respect the fact that we will obtain similar results for the gluelump mass is more than rewarding. 

Beyond this point, significant differences with the analysis of the previous section arises. Unlike $\delta m_L$, $V_A$ is infrared sensitive to ultrasoft effects. This generates logarithmic divergences that mix with genuine $\mathcal{O}(r^2)$ ultrasoft effects. The latter have been explicitly accounted for by introducing $\delta E^{(2)\PV}_{A,us}(r;\nu_{us})$ in Eq. (\ref{Method2}). The large logarithms generated by the factorization between the soft and ultrasoft scale are incorporated in $\delta V_A^{\rm RG}(r)$. They have been computed in Refs. \cite{Pineda:2000gza,Brambilla:2006wp,Brambilla:2009bi,Pineda:2011aw,Pineda:2011db}.

Overall, the precision of Eq. (\ref{Method2}) is $o( r^2)$. This is to be compared with the precision ${\cal O}(a^2)$ of Eq. (\ref{LambdabarPV}). Therefore, in principle, we now have a more precise expression. In practice this is not really so because  this would require knowing $\delta E^{(2)\PV}_{A,us}(r;\nu_{us})$. The knowledge of this object is limited, as it is debatable whether it can be computed using a weak coupling approximation. In our analysis we have included it in our error budget by computing it either assuming that can be approximated to its weak coupling expression or using a $Ar^2$ nonperturbative ansatz instead.  

By fitting Eq. (\ref{Method2}) against high-precision lattice data sets for short-distance hybrid potentials \cite{Bali:2003jq,Schlosser:2021wnr}, we obtain
\begin{equation}
\Lambda_{B}^{\rm PV} = 2.38(11) r_0^{-1}
\end{equation}
This result is perfectly flat over variations of $r$ within errors, as one can see in Figs. 10 and 11 in Ref. \cite{Ayala2025}. This confirms the expected scale independence of the result. The result is also perfectly consistent with our previous determination. This confirms the expected scheme independence of the result, as the present analysis have been made in the $\MS$ and the previous one in the lattice scheme. The agreement between these two determinations also yields a confirmation of the application of the nonperturbative OPE (i.e. of effective field theories) to these observables. 

%%%%%%%%%%%%%%%%%%%%%%%%%%%%%%%%%
\section{Conclusions and Future Outlook}
The extraction of the lightest gluelump mass via hyperasymptotic expansions marks a watershed moment in non-perturbative QCD. The two independent determinations are remarkably consistent and combine to yield \cite{Ayala2025}:
\begin{equation}
\Lambda_{B}^{\rm PV} = 2.44(7) r_0^{-1}
\end{equation}
This framework conclusively proves that even when probing the deep asymptotic regime of perturbative QCD, perturbation theory provides rigorously exact information, provided the divergent behavior of the perturbative series is correctly tamed. The extracted gluelump mass is fully renormalization-group invariant and independent of the renormalization scheme used for the strong coupling constant. On top of that, the outcome is free of the spurious power-like infrared dependencies that plagued earlier hard-cutoff renormalon subtraction schemes. These results provide an exceptionally robust foundation for future effective field theory studies of XYZ exotic hadrons and the spectrum of heavy quarkonium hybrids.

\bigskip

\noindent
{\bf Acknowledgments.}\\
I am deeply grateful to Cesar Ayala for the collaboration that lead to this research. This work was supported in part by the Spanish Ministry of Science and Innovation PID2023-146142NB-I00.

\end{document}